# Roland Eötvös: scientist, statesman, educator[1]

**András PATKÓS**
**Institute of Physics, Eötvös University**

**Abstract.** This lecture recalls the memory of Baron Roland Eötvös, an outstanding figure of the experimental exploration of the gravitational interaction and "funding father" of applied geophysics. Beyond the scientific achievements his contribution to the development of the modern Hungarian schooling and higher educational system, most importantly, the foundation of an innovative institution of teacher's training did not lose its contemporary significance. This lecture has been invited by the organizers of this Conference in response to the decision of UNESCO to commemorate worldwide the death centenary of the most outstanding Hungarian experimental physicist of modern times.

## 1. Curriculum vitae [1]

The noble family Eötvös has played an important role in the19th century history of Hungary. József (Joseph) Eötvös was a prominent political leader of the social reform period before the 1848 revolution and became minister of education and religious cults in the first independent Hungarian government. After the radicalisation of the revolution he has spent a short period in exile. Later, he has participated actively in the political negotiations leading to the formation of the Austro-Hungarian Monarchy in 1867, after which once again he entered the Hungarian government.

József Eötvös, though he has recognized early the affection of his son, Loránd (Roland) to investigating natural phenomena, still has convinced him to study for two years law at University of Pesth before joining University of Heidelberg. He wrote to his son in 1866: "If your soul, tired in sciences after a few years, would aim at activities on a wider horizon, I could not justify myself depriving you of the necessary basic legal education."

The academic and public carrier of Roland Eötvös took a very steep start after he has received the doctoral degree in Heidelberg as one can see from the Table next page. There a short list of the important events of his life for the curious reader is encompassed in the second column by his age. As a highly respected scientist he served in several positions at University of Budapest and has been elected as the ever youngest president of the Hungarian Academy of Sciences. He has initiated the foundation of the Mathematical and Physical Society, and of the Hungarian Tourist Association. He acted for a short time as Minister of Cults and Education. During his ministerial commitment he has finalized the laws ensuring religious freedom and has enlarged the network of the "elementary" schools. With the foundation of the József Eötvös College, and earlier of the Mathematical and Physical Society he has contributed innovatively at improving the formation and the working conditions of high-school teachers.

| Year | Age | Family/Private | Scientific&Academic | Public&Social |
|---|---|---|---|---|
| 1848 | | Born in *Buda* | | |
| 1868-70 | 20-22 | PhD Studies in *Heidelberg* | | |
| 1871 | 23 | | Habilitation at University of *Pesth* | |
| 1873 | 25 | | Corresponding member of Hungarian Academy of Sciences | Editor of the science popularization monthly Communications in Natural Sciences |
| 1876 | 28 | Married on Gizella Horvát (2 daughters) | **New method for the experimental determination of the *capillary constant*** | 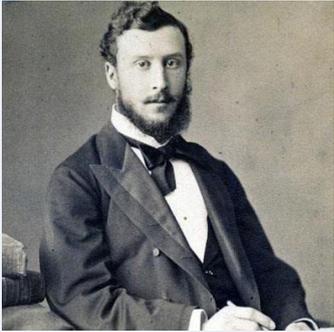 |
| 1878 | 30 | | Head of the Department for Experimental Nature Studies at University of *Budapest* | |
| 1880 | 32 | Yearly summer mountaineering till 1914 | | Vice-President of the Society for Advancement of Natural Sciences (till 1919) |
| 1886 | 38 | | **Eötvös-rule of the *temperature dependence* of the capillary constant** | |
| 1888-90 | 40-42 | 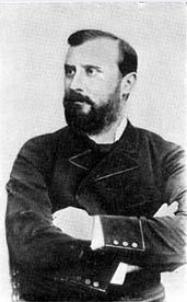 | **New method for *testing the law of Universal Free Fall*** | 1889: President of the Hungarian Academy of Sciences (till 1905) |
| 1891 | 43 | | **Construction of the *Eötvös-balance*** | **Foundation of the Mathematical and Physical Society (president till 1919)** |
| 1892-95 | 44-47 | | 1892-93: Rector of the University of Budapest | Minister of Education and Cults (1894-95) |
| 1895 | 47 | | | **Foundation of the József Eötvös Collegium (curator till 1919)** |
| 1906-1909 | 58-61 | 1902: Eötvös peak in Dolomites | ***Improved bound* on the validity of Universal Free Fall** | 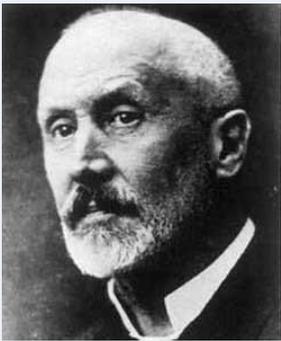 |
| 1913 | 65 | | **Laboratory demonstration of the *Eötvös-effect*** | |
| 1919 | 71 | Died in Budapest | | |

Note, that starting with the Paris meeting of the International Conference of Geodesy in 1900 he regularly has presented results obtained with the famous Eötvös balance and has received increasing international attention at the subsequent Budapest (1906) and Hamburg (1912) meetings. He has been nominated to the 1911 Nobel-prize (received eventually by Kammerlingh-Onnes).

On the occasion of his 70$^{th}$ birthday (1918) the Mathematical and Physical Society has published a volume reviewing his scientific and educational achievements, which has been reedited in 1930 by the Hungarian Academy of Sciences [2]. The Geophysical Institute founded by himself has taken his name directly after his death in 1921. Also, the Mathematical and Physical Society has been named after him (1923) as well as a College supporting outstanding students of poor social origin, originally initiated by the famous physicist Zoltán Bay, a former Eötvös-collegiate (1929). Since then many streets, schools, institutions bear his name in Hungary, including the University of Budapest (1950).

Below, in section 2 I shall discuss his scientific discoveries which preserve a place for his name in many monographies of chemistry, physics and geophysics. In Section 3 a brief summary is presented of his views and actions on improving the quality of high-school teaching.

## *2. Three important discoveries bearing the name of Roland Eötvös*

*2.1 Accurate experimental method for the determination of the capillary constant and the Eötvös-rule of its temperature dependence [3]*

In 1869 Eötvös has attended the lectures of Franz Neumann in Königsberg who also gave lectures on his theory of capillarity. The fluid wets and „climbs" up along the walls of a container as is shown in Fig.1 for the case of a liquid between two parallel glass plates. The height *($z_l$)* of a surface point *($I_l$)* measured relative to the horizontal surface of the fluid far from the wall is related to the angle between the external normal and the vertical directions *($\vartheta_l$)* of a ray of light reflected exactly horizontally at the point through a simple functional form containing the capillary constant *α* hidden in a constant *a* containing also the gravitational acceleration *g* and the density ρ of the liquid. The relation is determined by the equation proposed independently by Gauss, Laplace and others for the determination of the curvature radius *R* of the liquid surface:

$$\frac{1}{R} = \frac{2z}{a^2}, \quad a = \frac{2\alpha}{g\rho}.$$

Student Eötvös was praised by his professor for having proposed a method for measuring the capillary constant by determining the height difference of two surface points and the corresponding angles.

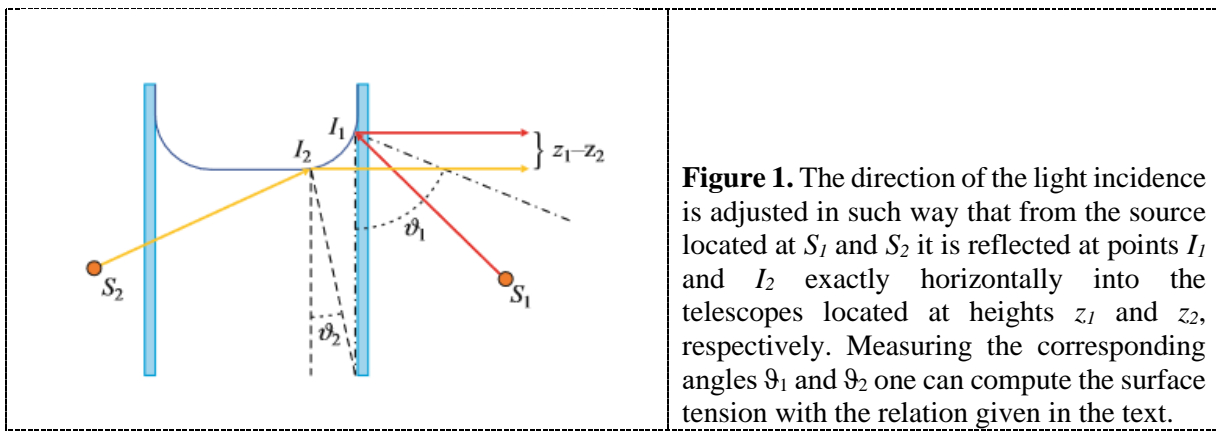

**Figure 1.** The direction of the light incidence is adjusted in such way that from the source located at $S_1$ and $S_2$ it is reflected at points $I_1$ and $I_2$ exactly horizontally into the telescopes located at heights $z_1$ and $z_2$, respectively. Measuring the corresponding angles $\vartheta_1$ and $\vartheta_2$ one can compute the surface tension with the relation given in the text.

Eötvös has solved the differential equation arising when one expresses the radius of curvature through the derivative of the *z(x)* function where *x* is the horizontal distance of the point *I* measured from the wall:

$$R = \frac{(1+z'^2)^{\frac{3}{2}}}{z''}, \quad z' = \frac{dz}{dx}.$$

From the solution $\frac{z^2}{a^2} = 2(\sin\frac{\vartheta}{2})^2$ one easily forms the convenient difference which allows the accurate determination of the combination *a*:

$$z_1 - z_2 = \sqrt{2}a(\sin\frac{\vartheta_1}{2} - \sin\frac{\vartheta_2}{2}).$$

Eötvös has published his method only in 1876. His superior method allowed him to pursue a rather extended investigation of the temperature dependence of the surface tension.

Following the derivation of the general gas law based on the hypothesis of molecular substructure by van der Waals, he has derived an analogous law for the surface tension (Eö is the *Eötvös-constant*):

$$\alpha V^{2/3} = E\ddot{o}\,(T_* - T).$$

The structural analogy with the formula *pV=RT* is obvious. The Eötvös-constant plays the same role as the Regnault-constant R.

His student of the epoch, Professor Károly Tangl wrote in 1930: „Whenever the theoretical result has been formulated, feverish laboratory work has been started for the verification of the rule. The windows of the old physics building of the university have remained light frequently in the late nights."Eötvös and his collaborators have checked the universal nature of his law on a large number of materials before he has presented his results in 1885 at a session of the Academy of Sciences. He has identified $T_*$ with the critical temperature, where the surface tension vanishes. This has been corrected later by Ramsay to lie somewhat lower.

An important trait of the scientist's character can be observed already in this early story. His deductions were always based on experimental facts, obtained with a procedure as perfect as the measurement methods of the epoch have allowed to reach. He did not publish intermediate "progress reports", only the final and best results were presented to the scientific community.

*2.2  Measurement of the local variation of the gravitational potential with the Eötvös balance and checking universal proportionality of the inertial and gravitational masses with unprecedented accuracy [4]*

The scientific challenge of his life has met Eötvös at age 40. This was the Definition I of the *Principia* of Newton which states the universal proportionality of the inertial mass ("resisting" to the accelerating force) and the gravitational mass (deduced from the weight of a body), which Newton claims to have checked by the most accurate pendulum measurements (see Fig.2).

| | |
|---|---|
| **DEFINITIO I.**<br>*Quantitas materiæ est mensura ejusdem orta ex illius densitate et magnitudine conjunctim.*<br><br>AER densitate duplicata, in spatio etiam duplicato, fit quadruplus; in triplicato sextuplus. Idem intellige de nive & pulveribus per compressionem vel liquefactionem condensatis. Et par est ratio corporum omnium, quæ per causas quascunque diversimode condensantur. Medii interea, si quod fuerit, interstitia partium libere pervadentis, hic nullam rationem habeo. Hanc autem quantitatem sub nomine corporis vel massæ in sequentibus passim intelligo. Innotescit ea per corporis cujusque pondus: Nam ponderi proportionalem esse reperi per experimenta pendulorum accuratissime instituta, uti posthac docebitur. | **Figure 2.** "*per experimenta pendulorum accuratissime instituta*" Newton has checked the universal proportionality of inertial and gravitational masses with 1:1000 accuracy. This fundamental statement constitutes Definition I of his *Philosophiae naturalis principia mathematica* |

In 1890, when Eötvös has presented results from the first series of experiments checking this proportionality with 1:20.000.000 accuracy, he gave the following natural motivation for performing the experiments: "Logics requires to verify the fundamental statement with the accuracy one can measure the weight of a body." By the middle of the XIXth century weights of bodies could have been measured with $1:10^6$ maximal error, though the universal nature of the proportionality has been checked with help of the so-called reversible pendulum developed by F. Bessel in 1830 only to $2:10^5$.

Eötvös has replaced the pendulum by the Cavendish-Coulomb torsion balance and reached 1:20.000.000 accuracy. The working principle of the equipment can be simply derived with help of Fig. 3a.

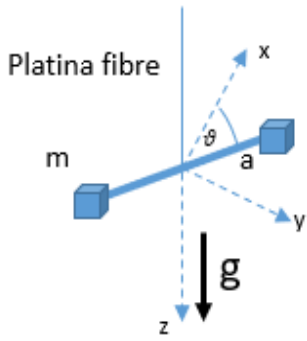
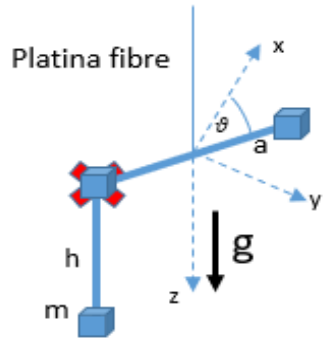

Fig.3a                              Fig.3b

One orients ideally the torsion fibre vertically (parallel to the local direction of the full gravitational acceleration in the point of hanging a beam of length *2a*, joining two pointlike bodies of mass *m* placed to the two ends). By the inhomogeneity of the gravitational potential the following force components act in the horizontal *(x,y)*-plane in the point *(Δx,Δy)* when the gradient of the gravitational potential energy is expanded to first order in the small coordinates:

$$F_x = \rho(0)(U_{xx}(0)\Delta x + U_{xy}(0)\Delta y), \quad F_y = \rho(0)(U_{yx}(0)\Delta x + U_{yy}(0)\Delta y).$$

Here $V(x) = \rho(x)U(x)$ is the gravitational potential energy density for a local mass density $\rho(x)$ and $U_{ij}$ is the second derivative of the gravitational potential with respect to the coordinates $x_i$ and $x_j$. Substituting these expressions into the formula giving the torque along the *z*-direction one finds:

$$M_z = \int (\Delta x F_y - \Delta y F_x) = (\theta_{xx} - \theta_{yy})U_{xy}(0) + \theta_{xy}(U_{yy}(0) - U_{xx}(0)),$$

where $\theta_{ij}$ are the components of the inertial momentum tensor. In the above geometry: $\theta_{xy} = ma^2 \sin\vartheta \cos\vartheta$, $\theta_{xx} - \theta_{yy} = ma^2[(\cos\vartheta)^2 - (\sin\vartheta)^2]$.

In 1891 Eötvös has designed a modified balance where the weight at one end is lowered by an amount $h$ (Fig. 3b). The advantage of this geometry is that it is sensitive also to two further components of the second derivative tensor of the gravitational potential. The additional torque is given as

$$\Delta M_z = a \cos\vartheta \, \Delta F_y - a \sin\vartheta \, \Delta F_x = ma\left[h\left(\cos\vartheta \frac{\partial g}{\partial y} - \sin\vartheta \frac{\partial g}{\partial x}\right)\right] = mah(U_{zy}\cos\vartheta - U_{zx}\sin\vartheta).$$

Now, there is a third contribution if the two masses of different material constitution experience different gravitational force, e.g. different values of the Newton's constant: $G' = G_0(1 + \kappa_1)$  $G = G_0(1 + \kappa_2)$, while the centrifugal force proportional to the inertial masses is identical for the two: $-maG_0 \sin\vartheta (\kappa_1 - \kappa_2) \sin\varepsilon$. Here $\varepsilon$ is the small angle between the direction of the weight vector and the (average) gravitational force. Its value is nearly 6° at the latitude of Budapest. The equation defining the change in the angular equilibrium direction of the beam relative to the torque-less case, $\varphi - \varphi_0$ has the following form ($\tau$ is the torsion constant of the fibre):

$$\tau(\varphi - \varphi_0) = K[U_{xy}(0)\cos 2\vartheta + (U_{yy}(0) - U_{xx}(0))\frac{\sin 2\vartheta}{2}] + mah(U_{zy}\cos\vartheta - U_{zx}\sin\vartheta) - maG_0 \sin\vartheta (\kappa_1 - \kappa_2) \sin\varepsilon.$$

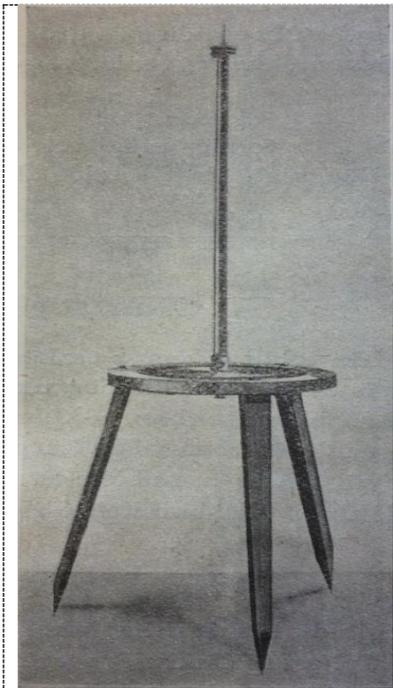

**Figure 4.** The Cavendish-Coulomb balance of Eötvös used in the 1888-90 series of experiments. Eötvös has improved its stability through careful selection of the torsion fibre and by thermal and aerial isolation.

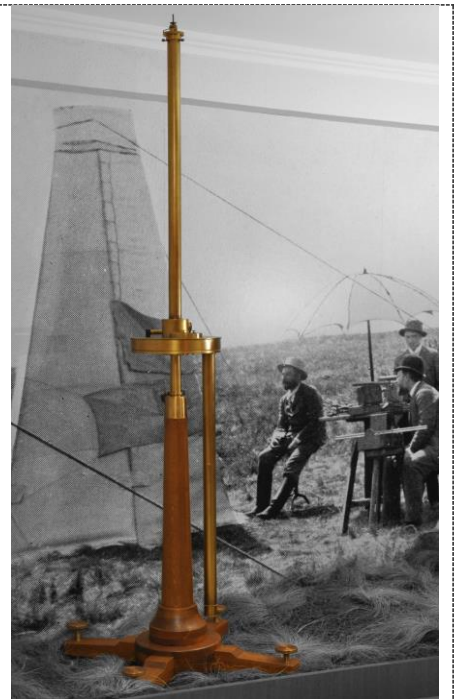

**Figure 5.** The first (asymmetric) Eötvös balance (1891), used in this picture in field experiments and gained importance in exploring oil resources in the USA, India Venezuela, etc. between 1920 and 1940. The balance was placed in the tent and observed with a theodolite.

In his first series of experiments Eötvös has used the simple Cavendish-Coulomb balance and has performed two series of measurements choosing the East-West direction for the beam orientation: $\vartheta =$

$\frac{\pi}{2}, \frac{3\pi}{2}$ (see Fig.6). He put first the same (platina) sample on both sides of the balance. In the second ("primed") series one of them has been replaced by a different (comparison) material. From the four equations one can express the difference $\kappa_{Pt} - \kappa_{Comparison}$ through the following simple formula

$$\tau(\varphi_{\pi/2} - \varphi_{3\pi/2}) - \tau(\varphi'_{\pi/2} - \varphi'_{3\pi/2}) = 2\tau(\kappa_{Pt} - \kappa_{Comparison})$$

New versions of the Eötvös-balance have allowed between 1906 and 1909 to improve the degree of the agreement of the inertial and gravitational masses further to 1:200.000.000.

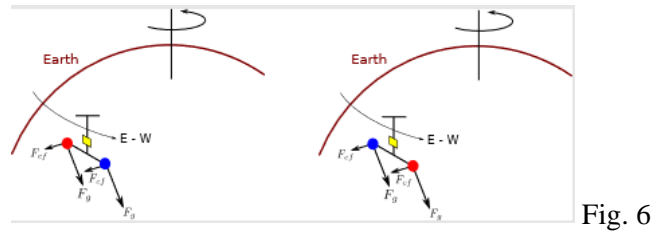
Fig. 6

The confirmation of the Equivalence Principle has been acknowledged with satisfaction by Einstein in January of 1918: "I cannot close my letter without expressing my gratitude for your work advancing greatly our knowledge on the identity of the gravitational and inertial masses."

Alternative gravitational theories have reinforced the significance of the Eötvös experiment. Robert Dicke and collaborators have performed in 1963 an Eötvös-type experiment in Princeton employing more advanced technology [5]. They compared the gravitational acceleration of various samples attracted by the Sun and improved the bound set by Eötvös by three orders of magnitude. Braginsky and Panov have performed their measurements in Moscow along the same concept in 1972. It is [6] interesting that in his famous (post-hume) paper dating 1922 Eötvös and his collaborators have also considered an experiment exploiting the gravitational attraction of the Sun. They recognized the great advantage of this setup, namely that the rotation of the Earth automatically exchanges the position of the samples relative to the Sun in every 24 hours. However, they have estimated lower sensitivity to this method compared to that in which the Earth's attraction is exploited and the samples are exchanged by hand.

In 1986 Ephraim Fischbach has reanalysed the original data of Eötvös and has discovered a non-trivial relation between the measured differences in the mass ratios and in the number of nucleons in 1 mol of the samples [7,8]. For the interpretation of this relation he has proposed the existence of an unknown new fundamental interaction ("Fifth Force") with a range of about 10-100 m. This conjecture has been checked by a research group led by Eric Adelberger at University of Seattle, again relying on an Eötvös-type balance (Fig.7). During the years this group has improved its torsion balance rotated with 1mHz angular velocity and by 2008 they could improve upon the original Eötvös results by five orders of magnitude [9]. However, no sign could be detected for the new hypothetical fundamental force.

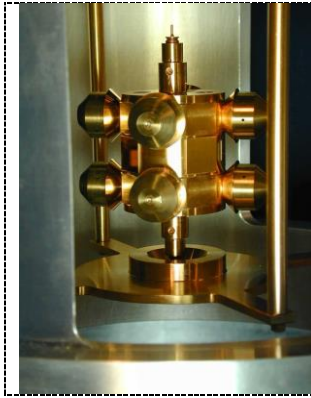

**Figure 7.** The Eötvös-type balance of the Eöt-Wash group (University of Washington, Seattle) rotates with 1mHz angular velocity. Its rotation period has been chosen to minimize the expected noise from the environmental oscillations and the instrument itself. The geographical location of the instrument has allowed to check the existence of any "fifth force" with longer than 1m range of action.

The search for such a new fundamental interaction hidden in the Newton-Einstein gravitational force is still in the forefront of actual researches. The latest tests were performed in the space in 2017, when a new experimental arrangement, different from the torsion balance has been employed [10]. Nevertheless, the first stage of these series of experiments has been started by Roland Eötvös about 130 years ago reserving for him a rightly deserved place in the high precision exploration of fundamental forces.

*2.3     Weight of a body in non-inertial motion relative to the Earth rotation [11]*

Between 1901 and 1905 Oskar Hecker (Potsdam Observatory) has studied the geographical variation of the gravitational acceleration on ships moving on the sea by comparing the height of the mercury column of a barometer with the atmospheric pressure obtained from the boiling point of water. In Hecker's data Eötvös has discovered the effect of the vertical component of the Coriolis force. Hecker has reconsidered the data as suggested by Eötvös. He has evaluated also new data obtained on two ships moving on the Black Sea in East-West, but opposite directions, and found perfectly consistent results for the gravitational acceleration. The modification of the effective gravitational acceleration due to the vertical component of the Coriolis-force is called *Eötvös effect*.

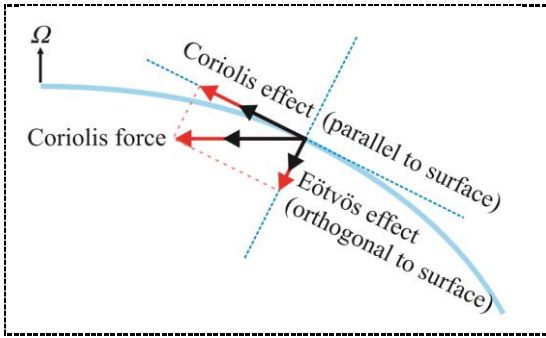

**Figure 8.** The Coriolis force for a body of mass $m$ moving with velocity $v$ eastwards at latitude φ (orthogonal to the plane of the graphics) has a component parallel to the surface deviating the motion of the body northwards. Eötvös called attention to the vertical component modifying the weight by $\Delta G = -2m\Omega \cos\vartheta v_{East}$, where $\Omega$ is the angular velocity of the Earth rotation.

Eötvös made use of the phenomenon of resonance for an indoor demonstration of the Eötvös effect. He constructed a seesaw on a rotating platform placing equal weights on the two ends of the beam (see Fig.9). The balance in the non-rotating state is in equilibrium, namely the two arms are in horizontal direction. When the platform starts to rotate one of the weights goes eastwards, while the other goes to west. In view of the Eötvös effect the weight of the two bodies changes in an opposite way. This is a very small effect. However, if the variation of the force acting on the ends follows the eigenfrequency of the seesaw then the amplitude of the seesaw will increase abruptly.

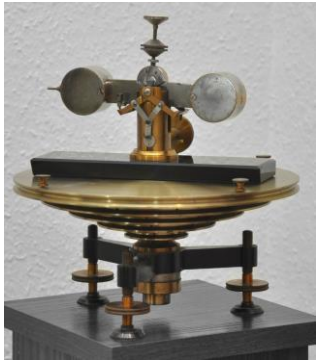 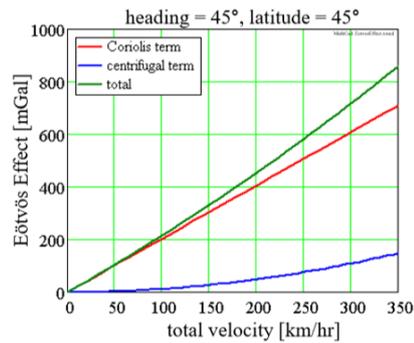

Fig. 9　　　　　　　　　　　　　　　　　　　　　　　　Fig. 10

The rotating balance of Eötvös represents an independent evidence for the Earth' rotation, equivalent to the famous pendulum experiment of Foucault. The discussion of the Eötvös correction (Fig.10) of airborne gravimetric measurements is found in all modern textbooks of geodesy.

## *3.   Initiatives for the modernization of teacher's training*

The completion of the works initiated by his father in the cultural and educational policy was a central ambition of Roland Eötvös during his short ministerial position. He has achieved the parliamentary approval of the last act of religious freedom, giving equal rights to the Jewish community with other religious groups. He has extended the network of elementary education by securing the finances for 400 new school buildings. He also understood the difficult financial situation of teachers and arranged in the state budget six times higher financial support for a foundation supporting teacher's families.

Here I wish to elaborate on his two most important initiatives which both exert important influence even today on the professional activity of the Hungarian teacher's community.

### *3.1   Foundation of the Mathematical and Physical Society (1891)*

After his nomination as ordinary professor of experimental physics (1878) he has participated very actively in the discussion on the modernization of public education and of higher education. In a dispute on the unification of the different forms of secondary education he has made the following statement: "The school needs a good school system and good teachers. You might ask which is more important. Myself, I would prefer the latter." He has added that in a suboptimal system the good teacher finds the way to teach well, but without good teachers even the best system will fail. Another time he has expressed his credo as it follows: *"The quality of teaching depends in first place on the scientific preparedness of the teachers."*

This sentence is a compressed expression of his "teacher-scientist" concept. According to Eötvös teachers should be able to follow the frontlines of science. He has proposed the foundation of the Mathematical and Physical Society as a medium where this sort of life-long self-education can be realized. Upon the foundation of the Society also its monthly periodical has been launched (Fig. 11).

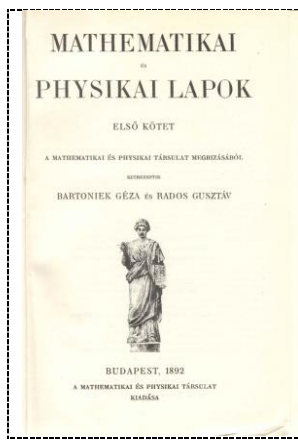
**Figure 11.** In the first issue of the Mathematical and Physical Journal Eötvös wrote in his salutation: "We shall write this journal for ourselves, not to publish original new results, but for presenting the latest results of the world science pedagogically, that every teacher could follow and make use of it in the teaching procedure."

The present journal of the Eötvös Physical Society, "Physics Review", has a separate section on Physics Teaching with its own editors, and publishes every month 2-3 papers, mainly by practising teachers and PhD-students developing innovative physics teaching methods and equipments for the demonstration of exciting natural phenomena.

*3.2   Foundation of the József Eötvös College (1895)*

This teacher's training college has been founded as an independent institution in form of a boarding-school (Fig.12). Its 100 members came from the countryside. 30 places were given to students from poor families without tuition fee payment obligation. The selection system of the new college members and the very strict requirements concerning the number of subjects to be studied and the performance level at the exams followed largely the norms of École Normale Supérieure (Paris). The College had from the very start one the best scientific libraries of the country and the students could attend language courses by lectors invited from countries where the given language is (one of) the official language(s).

Among the pupils of the nearly six decades when the College has followed closely the original prescriptions of Eötvös one can find the worldwide known composer Zoltán Kodály, but also the internationally esteemed mathematics teacher Tamás Varga or the legendary physics teacher Miklós Vermes who was the head of the jury of the Eötvös Physics Competition (1894) for half century.

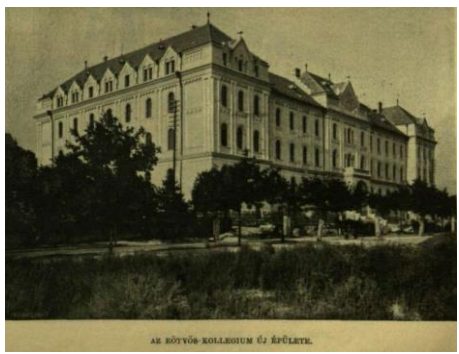
**Figure 12** The Building of the Eötvös College (1910)

The role of teachers is heavily disputed nowadays. Followers of some fashionable pedagogical directions consider the concept of Eötvös a kind of 19[th] century elitist ideal. But this ideal is still attractive: after the collapse of the "socialist" system which based its ideology on egalitarian pedagogical principles, new colleges were founded having in mind the example of the József Eötvös College. One of the first was the Bolyai College of the Faculty of Science of Eötvös University started in 1992.

A particularly important challenge concerns the way one should understand today the concept of "teacher-scientist". An initiative has been taken by the Hungarian Academy of Sciences in 2016, a program of Content Pedagogy development research. The goal is to propose ways to resolve hot problems of science teaching. Only project propositions with considerable participation of active high-school teachers could have chance to receive the 4 years support. Results obtained by the members of the Physics Education Research Group (one of the 19 supported research groups) will be presented in sections of this Conference. They can be considered Teacher-scientists of our days.

In summary of this part I wish to emphasize that the activities, the institutions and even the disputes demonstrate that the pedagogical ideas of Eötvös are still alive.

*4.   Conclusion*

This short lecture could not deal with all aspects of the colorful life of Roland Eötvös. There was no time and space to describe neither his science popularization activity nor his impressive alpine mountaineering record, documented through his stereographic photos made during his summer holidays in the Dolomites. I have put emphasis on his scientific achievements and on his initiatives improving the education and professional organization of physics teachers in Hungary. Telling his science story to our students should attract them towards the advances of modern gravitational physics. Introducing his character will positively impact on the personality of the young generation of teachers and researchers.

**Recommended literature:** A memorial volume of 14 essays covering many aspects of the life and works of Roland Eötvös was published [1] and is  freely downloadable from
*http://real-eod.mtak.hu/3799/1/001-176_tordelt1_ANGOL.pdf*